\documentclass[12pt,a4paper]{article}
\usepackage{fullpage}
\usepackage[centertags]{amsmath}
\usepackage{amsfonts}
\usepackage{amssymb}
\usepackage{eucal}
\usepackage{stmaryrd}
\usepackage[dvips]{graphicx,color}
\newcommand{\black}{\relax}
\newcommand{\blue}{\black}
\newcommand{\green}{\black}
\newcommand{\red}{\black}

\newcommand{\azzurro}{\black}

\newcommand{\porpora}{\black}
\newcommand{\rossoscuro}{\black}

\newcommand{\violascuro}{\black}
\begin{document}
\Large
\newcommand{\openone}{\leavevmode\hbox{\small1\kern-3.8pt\normalsize1}}
\newcommand{\slashed}[1]{ \ensuremath{ \rlap{\hskip0.5pt/} {#1} } }
\newcommand{\vet}[1]{\ensuremath{\hskip-1pt\vec{\hskip1pt#1}}}
\newcommand{\vett}[1]{\ensuremath{\hskip-0.5pt\vec{\hskip0.5pt#1}}}
{
  \LARGE
  \begin{center}
    \textbf{\porpora\LARGE{Statistical Analysis of Solar Neutrino Data}}
    \\[1cm]
    \textsf{\textbf{\violascuro{M.V. Garzelli}}}
    and
    \textsf{\textbf{\violascuro{C. Giunti}}}
    \\[1cm]
    {\azzurro{INFN, Sezione di Torino}}
    \\
    and
    \\
    {\azzurro{Dipartimento di Fisica Teorica,}}
    \\
    {\azzurro{Universit\`a di Torino}}
    \\
    {\azzurro{Via P. Giuria 1, I-10125 Torino, Italy}}
    \\[1cm]
    \texttt{\textbf{\rossoscuro{garzelli@to.infn.it}}}
    \\
    \texttt{\textbf{\rossoscuro{giunti@to.infn.it}}}
  \end{center}
}
\newpage 

\begin{equation*}
\text{\Large\textbf{\porpora{Standard Method}}}
\end{equation*}

{\green{Least-Squares}} estimator of
{\blue{$\Delta{m}^2$}} and {\blue{$\theta$}}:
{\red{$X^2_{\mathrm{min}}$}}
\begin{equation*}
X^2
=
\sum_{j_1,j_2}
\left( R^{\mathrm{(thr)}}_{j_1} - R^{\mathrm{(exp)}}_{j_1} \right)
(V^{-1})_{j_1j_2}
\left( R^{\mathrm{(thr)}}_{j_2} - R^{\mathrm{(exp)}}_{j_2} \right)
\end{equation*}
$R^{\mathrm{(thr)}}_{j}$
=
theoretical rate for experiment or bin $j$
\\
$R^{\mathrm{(exp)}}_{j}$
=
rate measured in experiment or bin $j$
\\
$j=1,\ldots,N_{\mathrm{exp}}$,
where $N_{\mathrm{exp}}$
is the number of data points
\\
$V$
=
covariance matrix: experimental and theoretical uncertainties
\begin{equation*}
\text{\porpora\textbf{Standard Goodness of Fit}}
\end{equation*}
Probability to observe a minimum
of $X^2$
larger than the one actually observed,
assuming for $X^2_{\mathrm{min}}$
a $\chi^2$ distribution with
$N_{\mathrm{dof}}=N_{\mathrm{exp}}-N_{\mathrm{par}}$
degrees of freedom
($N_{\mathrm{par}}$ is the number of fitted parameters).
\begin{equation*}
\text{\porpora\textbf{Standard Allowed Regions}}
\end{equation*}
The standard $100\beta\%$ CL regions
in the
$\tan^2\vartheta$--$\Delta{m}^2$ plane
are given by the condition
\begin{equation*}
X^2 \leq X^2_{\mathrm{min}} + \Delta{X^2}(\beta)
\end{equation*}
$\beta$ = Confidence Level (CL)
\\
$\Delta{X^2}(\beta)$
=
value of $\chi^2$ such that
the cumulative $\chi^2$ distribution for 2
degrees of freedom
is equal to $\beta$
\begin{align*}
\null & \null
\text{
$\beta=90\%$ ($ 1.64 \, \sigma$)
{\red{$\Rightarrow$}}
$\Delta{X^2}(0.90) = 4.61$
}
\\
\null & \null
\text{
$\beta=99\%$ ($ 2.58 \, \sigma$)
{\red{$\Rightarrow$}}
$\Delta{X^2}(0.99) = 9.21$
}
\end{align*}

\newpage 

\begin{equation*}
\text{\underline{\textsf{Analysis of Rates}}}
\end{equation*}
\begin{equation*}
\parbox{0.9\textwidth}{
Rates of Homestake \cite{Homestake-98},
GALLEX+SAGE \cite{GALLEX-99,SAGE-99},
Super-Kamiokande 2001 \cite{SK-sun-01}
${\red{\Rightarrow}}$
$N_{\mathrm{exp}}=3$,
$N_{\mathrm{par}}=2$,
$N_{\mathrm{dof}}=1$.
}
\end{equation*}

\begin{equation*}
\text{\underline{\textsf{Global Analysis}}}
\end{equation*}
\begin{equation*}
\parbox{0.9\textwidth}{
Rates of Homestake \cite{Homestake-98},
GALLEX+SAGE \cite{GALLEX-99,SAGE-99},
Super-Kamiokande 2001 \cite{SK-sun-01}:
3 data points
}
\end{equation*}
\begin{equation*}
+
\end{equation*}
\begin{equation*}
\parbox{0.9\textwidth}{
Shape of Super-Kamiokande 2000 \cite{SK-sun-00} Day-Night data:
6 bins and 1 normalization factor
}
\end{equation*}
\begin{equation*}
+
\end{equation*}
\begin{equation*}
\parbox{0.9\textwidth}{
Shape of Super-Kamiokande 2001 \cite{SK-sun-01} energy spectrum
for $E > 5.5 \, \mathrm{MeV}$:
18 bins and 1 normalization factor
}
\end{equation*}
\begin{equation*}
{\red{\Downarrow}}
\end{equation*}
\begin{equation*}
\text{
$N_{\mathrm{exp}}=27$,
$N_{\mathrm{par}}=4$,
$N_{\mathrm{dof}}=23$.
}
\end{equation*}

\begin{equation*}
\fbox{$ \displaystyle
\begin{array}{l}
\text{\textsf{Active}}
\ {\red{\Rightarrow}} \
\nu_e\to\nu_{\mu,\tau}
\\
\text{\textsf{Sterile}}
\ {\red{\Rightarrow}} \
\nu_e\to\nu_{s}
\end{array}
$}
\end{equation*}

\newpage 

\includegraphics[bb=120 505 370 780, width=0.49\textwidth]{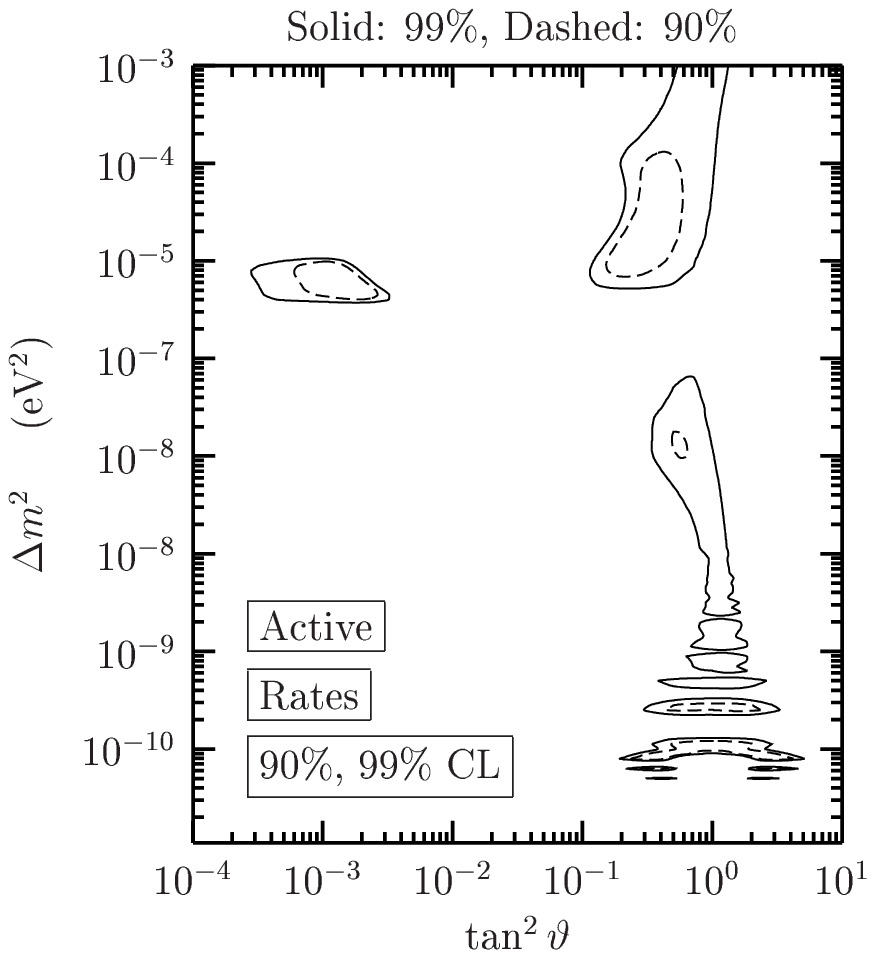}
\includegraphics[bb=120 505 370 780, width=0.49\textwidth]{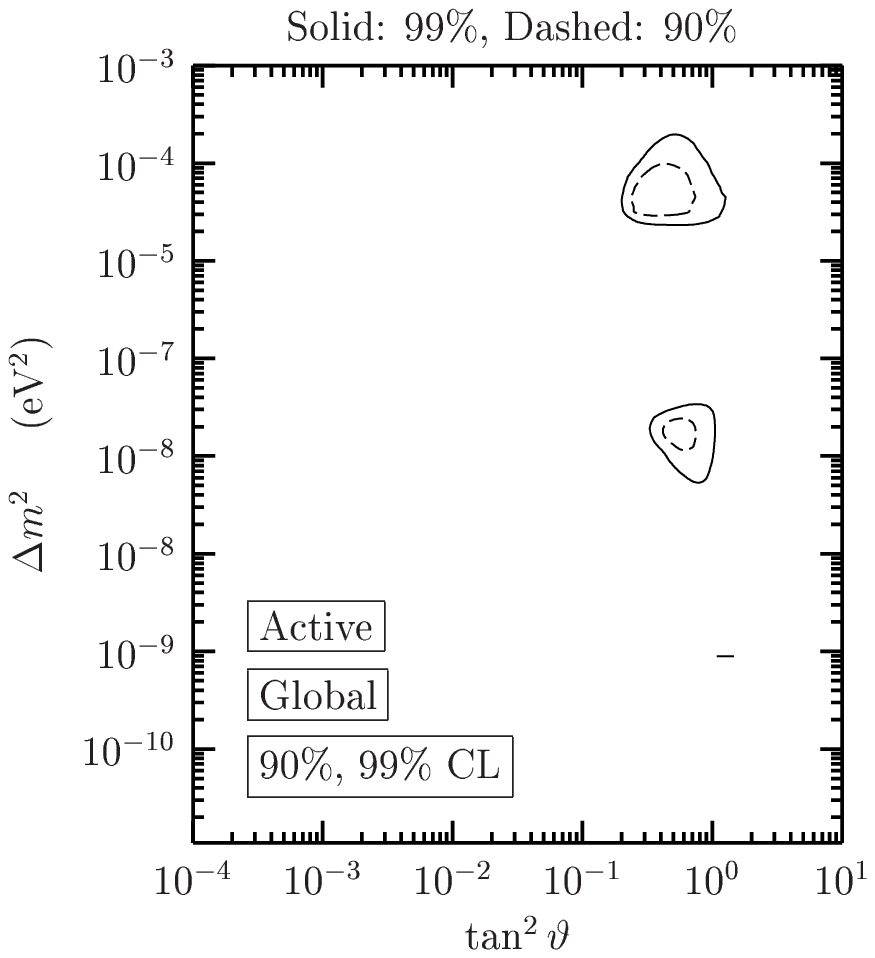}

\includegraphics[bb=120 505 370 780, width=0.49\textwidth]{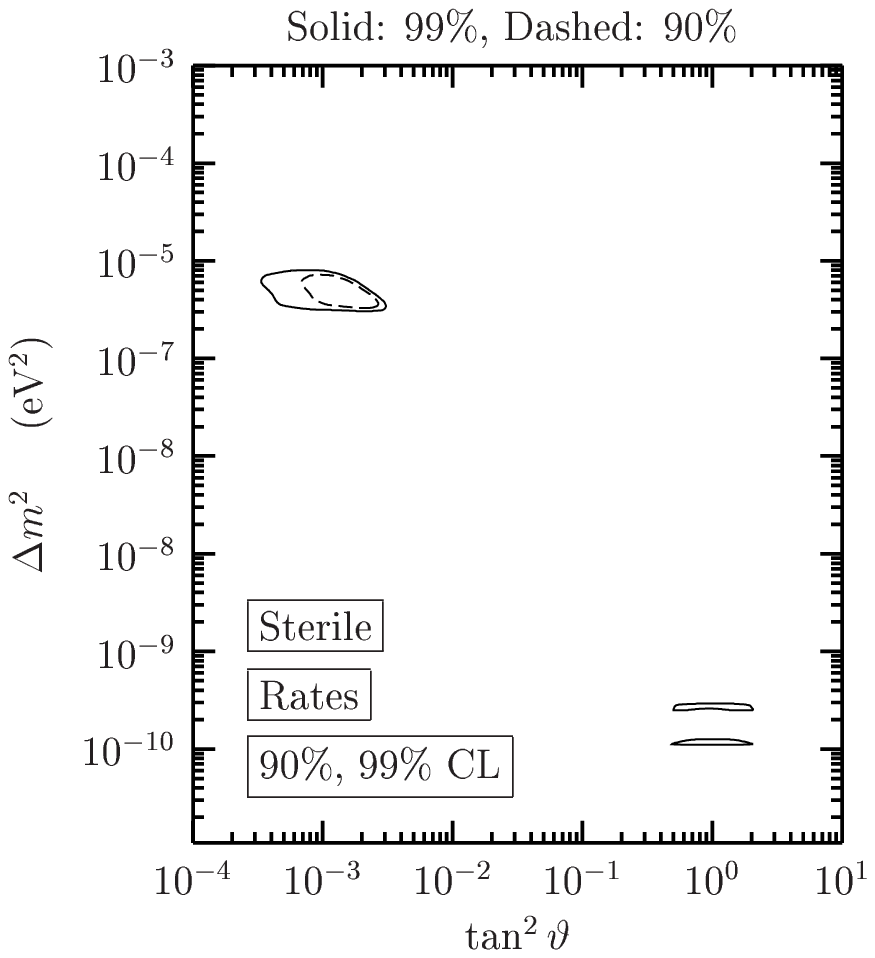}
\includegraphics[bb=120 505 370 780, width=0.49\textwidth]{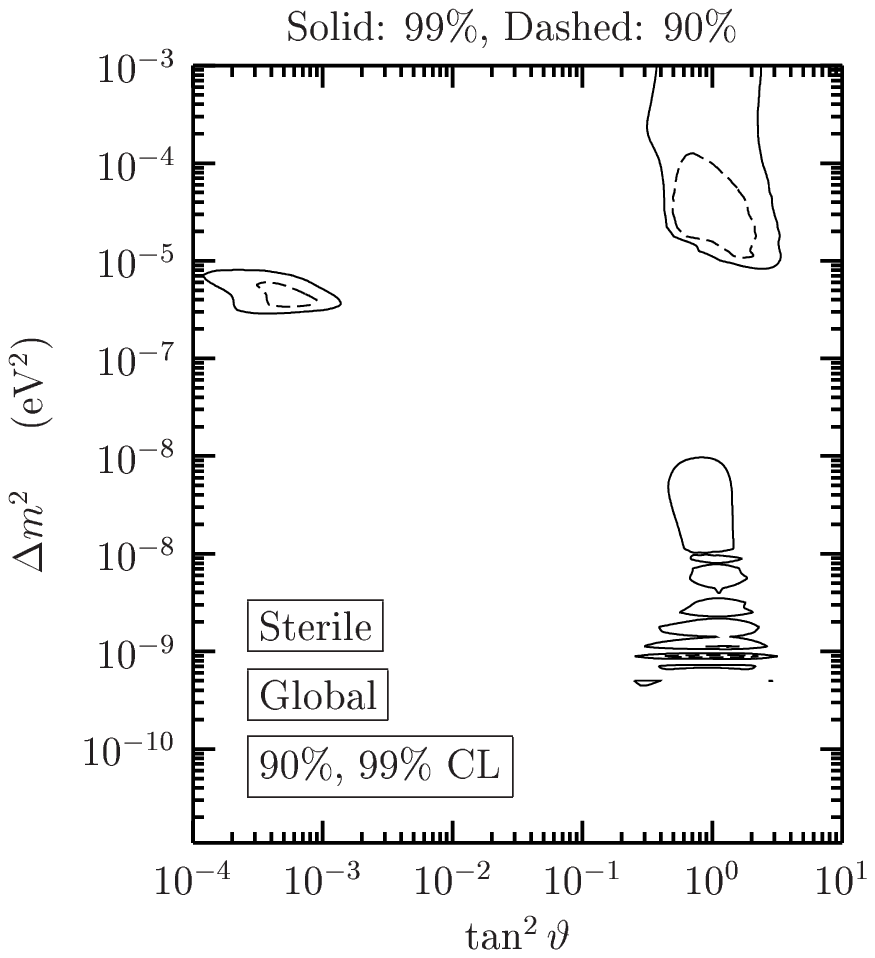}

\begin{equation*}
\text{Best fit values in Table at pag.~\pageref{tab-gof}.}
\end{equation*}

\newpage 

\begin{equation*}
\text{\textbf{\textsf{NOVELTIES}}}
\end{equation*}
Super-Kamiokande rate/BP-SSM has decreased from
$0.474 \pm 0.020$ \cite{SK-sun-spectrum-98}
to
$0.451 \pm 0.008$ \cite{SK-sun-01}.
%
\begin{equation*}
\text{\underline{\textsf{Active Rates}}}
\end{equation*}
\textbf{Best Fit in VO region!}
Due to decrease of SK rate/BP-SSM.
\begin{equation*}
\text{\underline{\textsf{Active Global}}}
\end{equation*}
\textbf{Best Fit continues to be in LMA region!}
\\
\textbf{No SMA region at 99\% CL!}
Due to flat spectrum.
\\
\textbf{At 99\% CL VO region almost vanish!}
Due to flat energy spectrum and Day-Night asymmetry
(albeit small).
\begin{equation*}
\text{\underline{\textsf{Sterile Rates}}}
\end{equation*}
\textbf{Best Fit continues to be in SMA region!}
\begin{equation*}
\text{\underline{\textsf{Sterile Global}}}
\end{equation*}
\textbf{Poor Goodness of Fit $\Longrightarrow$ sterile disfavored!}
\\
\textbf{Best Fit in LMA region!}
Due to flat energy spectrum
and decrease of Super-Kamiokande rate/BP-SSM
(the incompatibility of a flat energy spectrum
with the Homestake rate is alleviated).

\newpage 

\begin{equation*}
\text{\textbf{\textsf{Conditions for the validity of the Standard Method}}}
\end{equation*}
\fbox{\textbf{1}}
The theoretical rates
$R^{\mathrm{(thr)}}_{j}$
depend \underline{linearly} on the parameters
$\Delta{m}^2$ and $\tan^2\!\theta$
to be determined in the fit.
\\[0.3cm]
\fbox{\textbf{2}}
The errors
$R^{\mathrm{(thr)}}_{j}-R^{\mathrm{(exp)}}_{j}$
are \underline{multinormally} distributed.
\\[0.3cm]
\fbox{\textbf{3}}
The covariance matrix $V$
does not depend on $\Delta{m}^2$ and $\tan^2\!\theta$.

In reality these three conditions are \underline{not} satisfied:
\\[0.3cm]
\fbox{\textbf{1}}
The theoretical rates
$R^{\mathrm{(thr)}}_{j}$
do not depend at all linearly on the parameters
$\Delta{m}^2$, $\tan^2\!\theta$.
This is the reason why there are several allowed regions
in the
$\tan^2\!\theta$--$\Delta{m}^2$
plane
and these regions do not have elliptic form.
\\[0.3cm]
\fbox{\textbf{2}}
The errors
$R^{\mathrm{(thr)}}_{j}-R^{\mathrm{(exp)}}_{j}$
are not multinormally distributed,
because although the fluxes $\phi_i^{\mathrm{SSM}}$
and the cross sections $C_{ij}^{\mathrm{(thr)}}$
are assumed to be multinormally distributed,
their products,
that determine the theoretical rates
through the relations
\begin{equation}
R^{\mathrm{(thr)}}_{j}
=
\sum_i C_{ij}^{\mathrm{(thr)}} \, \phi_i^{\mathrm{SSM}} 
\,,
\label{Rj}
\end{equation}
are not multinormally distributed.
\\[0.3cm]
\fbox{\textbf{3}}
The covariance matrix $V$ depends on $\Delta{m}^2$ and $\tan^2\!\theta$.

\newpage 

\fbox{\textbf{1}}
is \underline{important}!

\fbox{\textbf{2}}
is \underline{irrelevant}:
the multinormal approximation is very good,
as shown by the following three figures.

\textbf{Dotted lines}: probability distribution function of
experimental rates generated with Monte Carlo.
\\
\textbf{Solid lines}: probability distribution function of
experimental rates assuming a multinormal distribution
given by the Likelihood function
\begin{equation*}
\mathcal{L}(R^{\mathrm{(exp)}}_{j}|\tan^2\!\theta,\Delta{m}^2)
=
\frac{e^{-X^2/2}}{(2\pi)^{N_{\mathrm{exp}}/2}\sqrt{|V|}}
\end{equation*}

\includegraphics[bb=80 505 520 770, width=\textwidth]{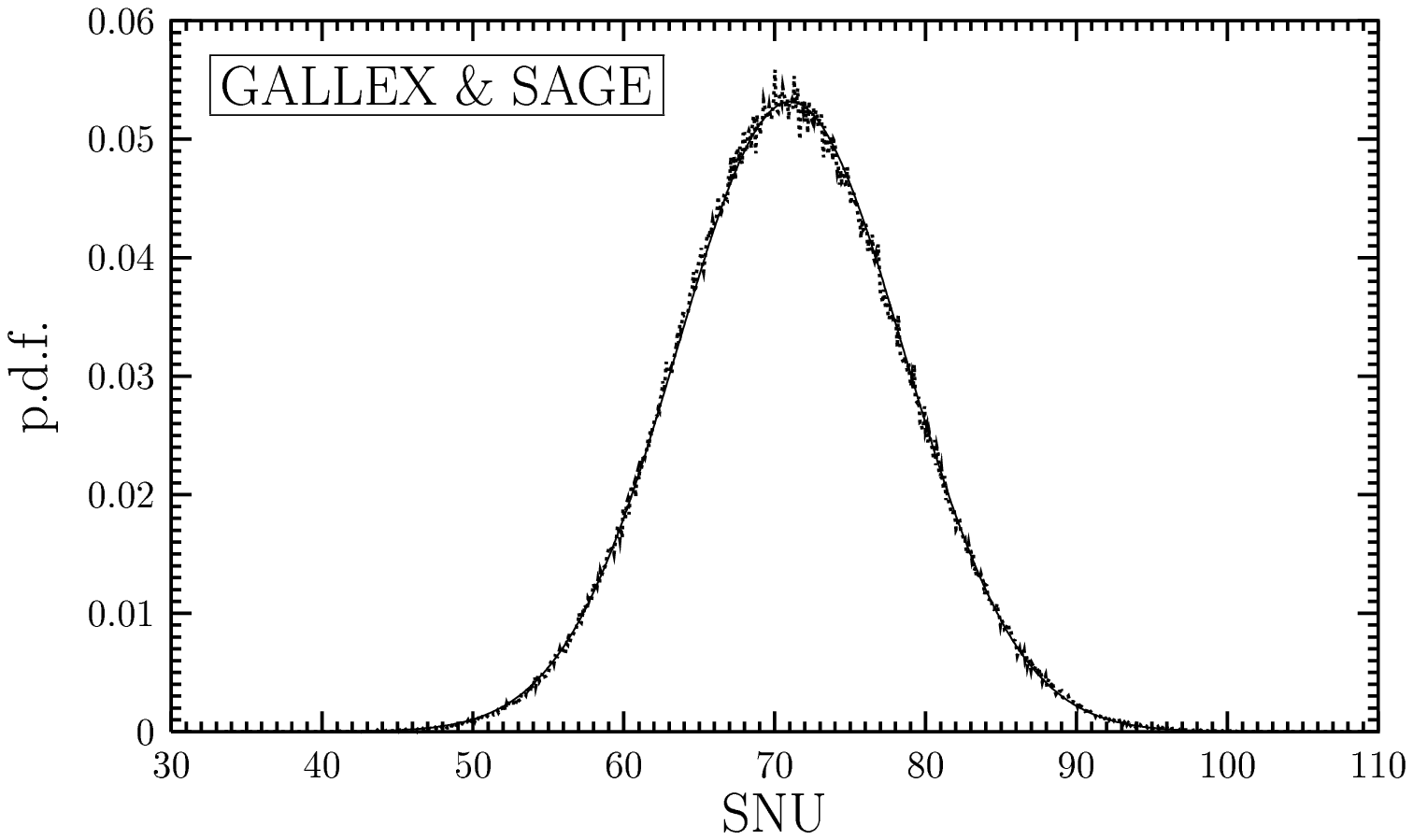}
\includegraphics[bb=80 505 520 770, width=\textwidth]{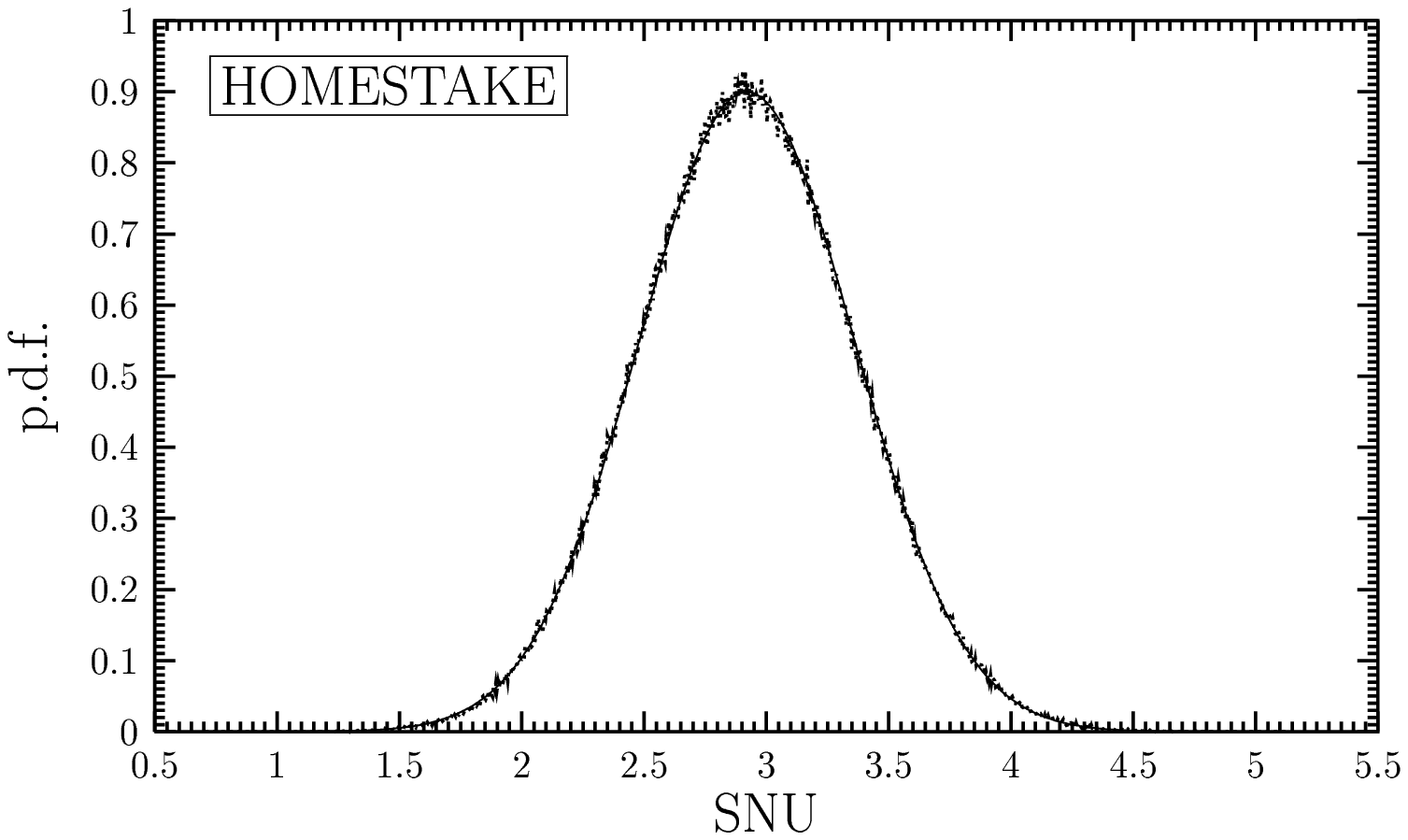}
\includegraphics[bb=80 505 520 770, width=\textwidth]{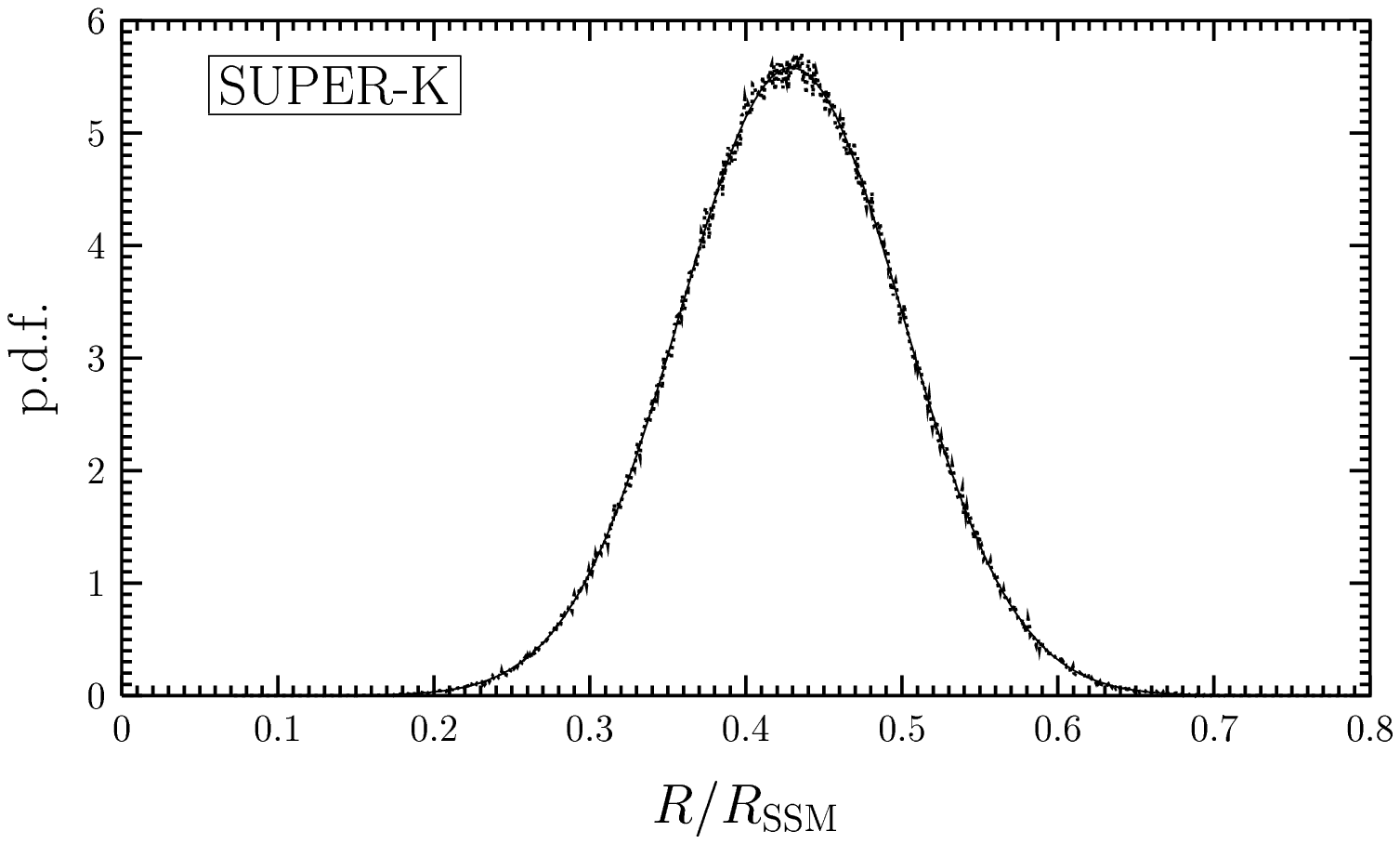}

\newpage 

\fbox{\textbf{3}}
is \underline{not negligible},
as shown by the following comparison of the Standard Regions
with the regions obtained with the log-Likelihood method (see \cite{Eadie-71})
\begin{equation*}
\ln\mathcal{L}
\geq
\ln\mathcal{L}_{\mathrm{max}} - \frac{\Delta{X^2}(\beta)}{2}
\end{equation*}
\includegraphics[bb=120 505 370 780, width=0.49\textwidth]{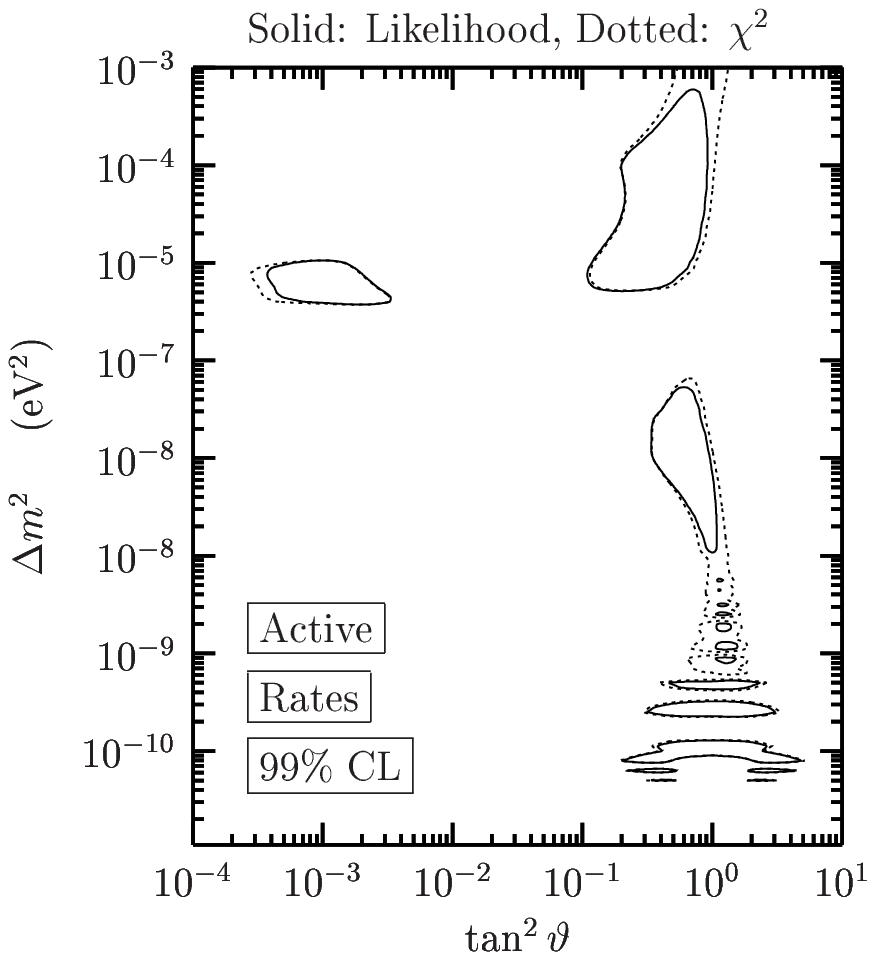}
\includegraphics[bb=120 505 370 780, width=0.49\textwidth]{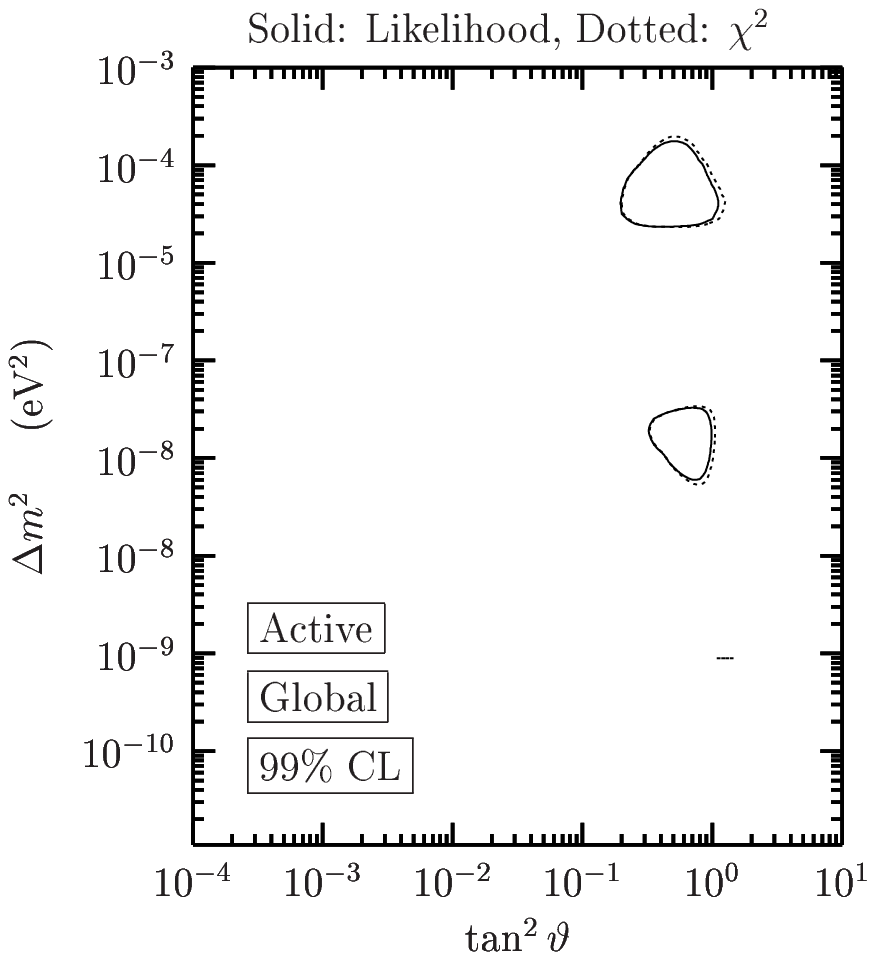}

\includegraphics[bb=120 505 370 780, width=0.49\textwidth]{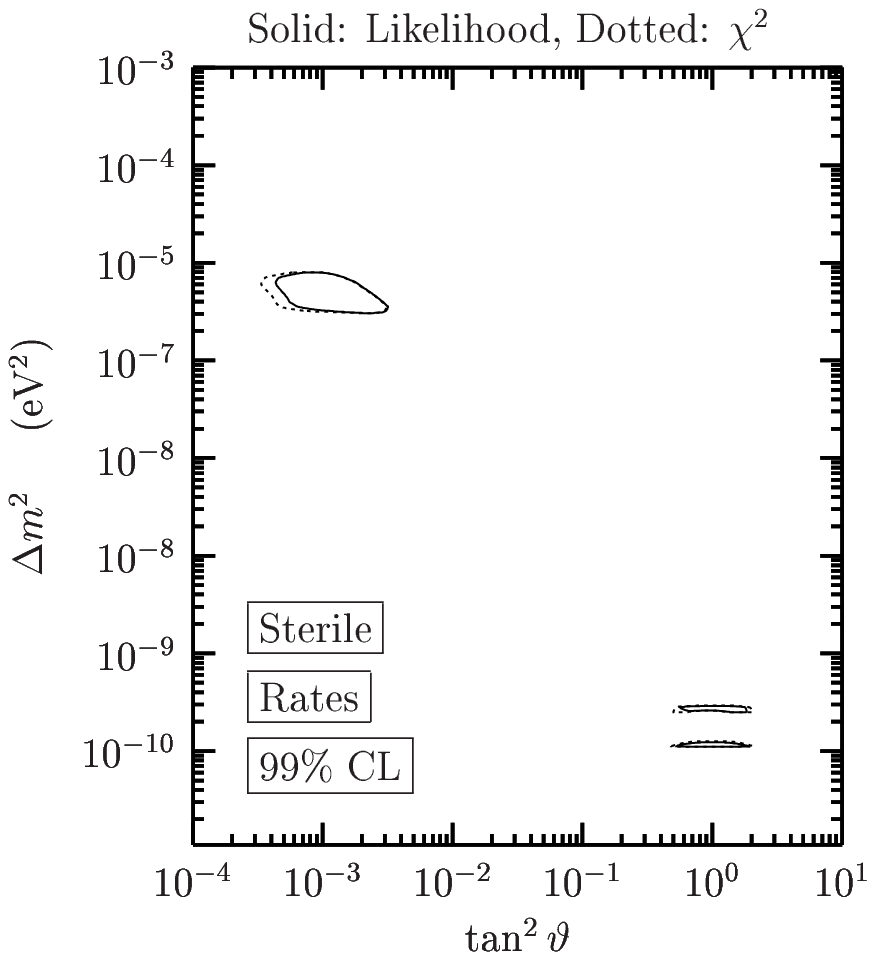}
\includegraphics[bb=120 505 370 780, width=0.49\textwidth]{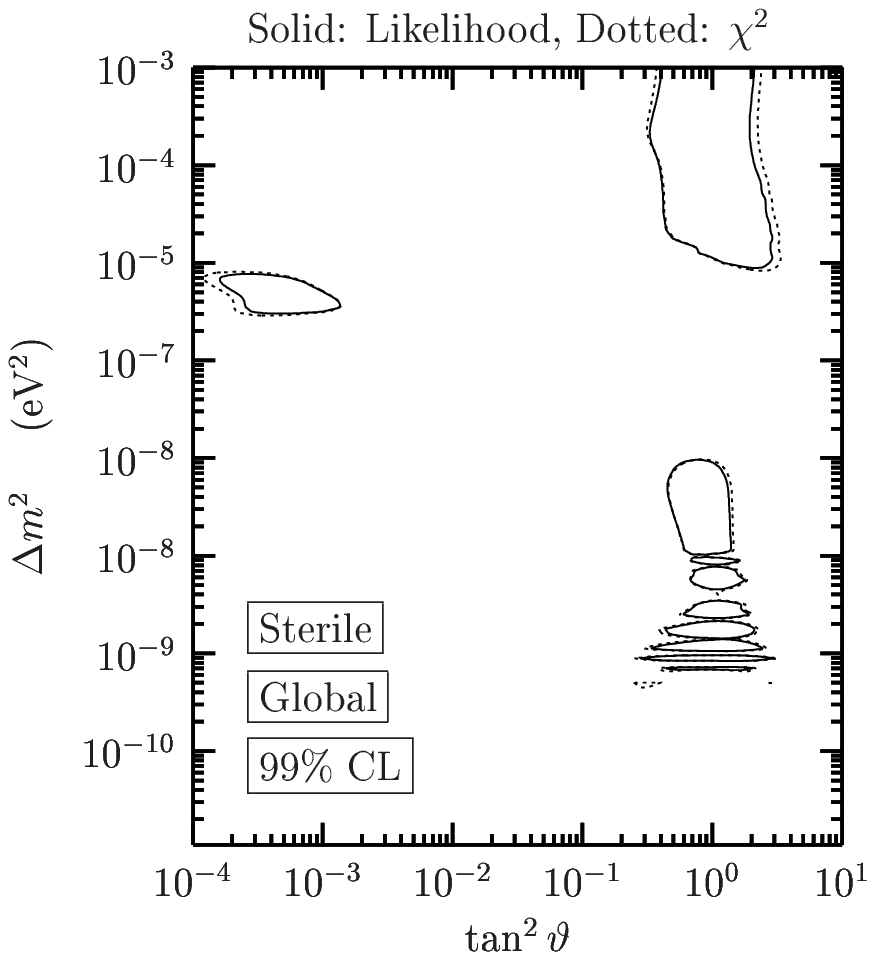}

\newpage 

\begin{equation*}
\text{\Large\textbf{\porpora{Monte Carlo Goodness of Fit
\protect\cite{Garzelli-Giunti-sf-00}}}}
\end{equation*}

{\large\red{$\blacktriangleright$}}
Estimate best-fit values of
$\Delta{m}^2$, $\tan^2\!\theta$
through $X^2_{\mathrm{min}}$.
\\
{\large\red{$\blacktriangleright$}}
Call the best-fit values
$\widehat{\Delta{m}^2}$, $\widehat{\tan^2\!\theta}$.
\\
{\large\red{$\blacktriangleright$}}
Assume that
$\widehat{\Delta{m}^2}$, $\widehat{\tan^2\!\theta}$
are reasonable surrogates of the true values
$\Delta{m}^2_{\mathrm{true}}$, $\tan^2\!\theta_{\mathrm{true}}$.
\\
{\large\red{$\blacktriangleright$}}
Using $\widehat{\Delta{m}^2}$, $\widehat{\tan^2\!\theta}$,
generate $N_s$ synthetic random data sets
with the standard gaussian distribution for the
experimental and theoretical uncertainties.
\\
{\large\red{$\blacktriangleright$}}
Apply the Least-Squares method
to each synthetic data set,
leading to an ensemble of simulated best-fit parameters
$\widehat{\Delta{m}^2}_{(s)}$, $\widehat{\tan^2\!\theta}_{(s)}$
with $s=1,\ldots,N_s$,
each one with his associated
$(X^2_{\mathrm{min}})_{s}$.
\\
{\large\red{$\blacktriangleright$}}
Calculate GoF
as the fraction of simulated
$(X^2_{\mathrm{min}})_{s}$
in the ensemble that are larger than the one actually observed,
$X^2_{\mathrm{min}}$.

\newpage 

\includegraphics[bb=87 539 420 773, width=\textwidth]{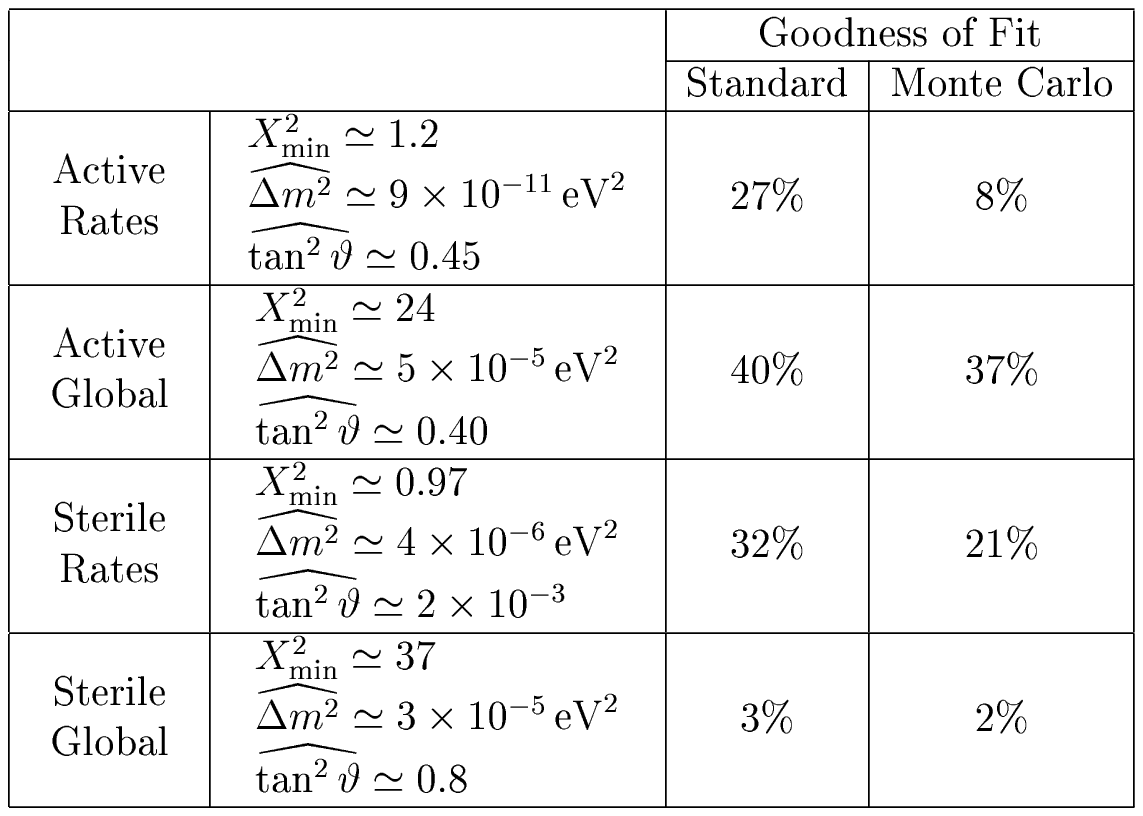}

\label{tab-gof}

\bigskip

{\Huge\red{$\rightsquigarrow$}}
\fbox{
Standard method \underline{overestimates} GoF!
}

\medskip

{\Huge\red{$\rightsquigarrow$}}
\fbox{\parbox{0.8\textwidth}{
\textbf{Explanation}:
GoF is the probability to obtain
better fits than the observed one.

When there are several local minima of $X^2$
with relatively close values of $X^2$,
there are more possibilities to obtain good fits
and the true goodness of fit is smaller
than the standard one
(obtained assuming that $X^2$
is a quadratic function of
$\tan^2\vartheta$, $\Delta{m}^2$,
with only one minimum).
}}

\newpage 

\begin{equation*}
\begin{array}{c}
\text{\Large\textbf{\porpora{Monte Carlo CL of}}}
\\
\text{\Large\textbf{\porpora{Standard Allowed Regions
\protect\cite{Garzelli-Giunti-sf-00}}}}
\end{array}
\end{equation*}

\textsf{\textbf{Definition}:}
$100\beta\%$ CL Allowed Regions
belong to a
set of allowed regions
that
cover the true value of the parameters
with probability $\beta$.

{\large\red{$\lightning$}}
Given the usual ``$100\beta\%$ CL'' allowed regions
in the $\tan^2\vartheta$--$\Delta{m}^2$ plane,
calculate their Monte Carlo Confidence Level
$\beta_{\mathrm{MC}}$
with a method similar to the one used
for the Goodness of Fit.

{\large\red{$\blacktriangleright$}}
Assume that
$\widehat{\Delta{m}^2}$, $\widehat{\tan^2\!\theta}$
are reasonable surrogates of the true values
$\Delta{m}^2_{\mathrm{true}}$, $\tan^2\!\theta_{\mathrm{true}}$.

{\large\red{$\blacktriangleright$}}
Generate a large number of synthetic data sets.

{\large\red{$\blacktriangleright$}}
Apply the standard procedure to each synthetic data set
and obtain the corresponding ``$100\beta\%$ CL'' Standard Allowed Regions
in the $\tan^2\!\theta$--$\Delta{m}^2$ plane.

{\large\red{$\blacktriangleright$}}
Count the number of synthetic
``$100\beta\%$ CL'' Standard Allowed Regions
that cover the assumed surrogate
$\widehat{\Delta{m}^2}$, $\widehat{\tan^2\!\theta}$
of the true values.

{\large\red{$\blacktriangleright$}}
The ratio of this number and the total number of synthetically
generated data set gives the Confidence Level
$\beta_{\mathrm{MC}}$
of the ``$100\beta\%$ CL'' Standard Allowed Regions.

\includegraphics[bb=87 622 320 773, width=\textwidth]{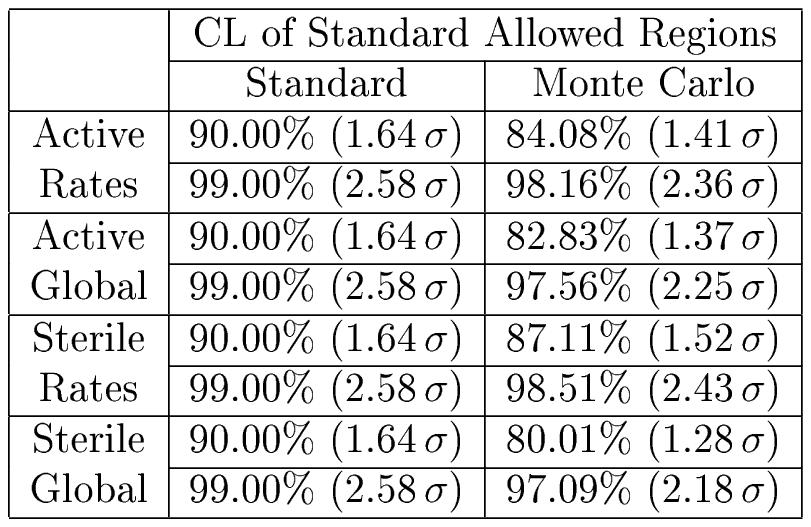}

\bigskip

{\Huge\red{$\rightsquigarrow$}}
\fbox{\parbox{0.8\textwidth}{
The Confidence Level of Standard Allowed Regions
is \underline{smaller} than its nominal value.
}}

\medskip

{\Huge\red{$\rightsquigarrow$}}
\fbox{\parbox{0.8\textwidth}{
\textbf{Explanation}: When there are several local minima of $X^2$
with relatively close values of $X^2$,
in repeated experiments
the global minimum has significant chances
to occur far from the true
value of $\tan^2\vartheta$ and $\Delta{m}^2$
${\red{\Rightarrow}}$
smaller probability that the allowed regions
cover the true value of $\tan^2\vartheta$ and $\Delta{m}^2$,
with respect to the case in which there is only one minimum.
}}

\newpage 

\begin{equation*}
\text{\Large\textbf{\porpora{Frequentist Allowed Regions}}}
\end{equation*}

{\large\red{$\Rightarrow$}}
\textbf{Frequentist Statistics}
allows to calculate allowed regions with correct \textbf{coverage}
using Neyman's method.

{\large\red{$\nRightarrow$}}
But there is arbitrariness in the choice of
\\
\phantom{\large\red{$\nRightarrow$}}
1) Estimator of the parameters
\\
\phantom{\large\red{$\nRightarrow$}}
2)
Method for the construction of acceptance regions

{\large\red{$\lightning$}}
In \cite{Garzelli-Giunti-sf-00}
we have calculated ``exact'' confidence regions
using as estimate of $\tan^2\vartheta$ and $\Delta{m}^2$
their value at $X^2_{\mathrm{min}}$.

{\large\red{$\curlywedgedownarrow$}}
In \cite{Creminelli-Signorelli-Strumia-01}
it has been argued that $X^2_{\mathrm{min}}$
may be an insufficient estimator,
leading to a loss of information.
Notice that if this is true,
the standard $\chi^2$ method suffers from the same problem!

{\large\red{$\curlywedgedownarrow$}}
In order to prevent any loss of information,
it is better to use the full data set
as estimator of $\tan^2\vartheta$ and $\Delta{m}^2$,
as done in \cite{Creminelli-Signorelli-Strumia-01}.

{\large\red{$\nRightarrow$}}
However,
there is still the problem of choice of the method
for the construction of acceptance intervals.

{\large\red{$\curlywedgedownarrow$}}
In \cite{Creminelli-Signorelli-Strumia-01}
it has been argued that the Unified Approach (UA) \cite{Feldman-Cousins-98}
is more appropriate than the smallest
acceptance intervals method,
also known as ``Crow--Gardner'' (CG).

\newpage 

{\large\red{$\nRightarrow$}}
Unfortunately,
it is well known that
when the UA differs from
the smallest
acceptance intervals method
it gives unreliable confidence intervals
(see \cite{Giunti-Laveder-physical-00,Giunti-Laveder-power-00})

{\large\red{$\nRightarrow$}}
Infamous example:
\\
The \textbf{KARMEN} 1998 limit on $\bar\nu_\mu\to\bar\nu_e$ oscillations
obtained with the Unified Approach
was unreliably much more stringent than the sensitivity of the experiment
\cite{KARMEN-nu98}.
\\
The \textbf{KARMEN} 1999 limit is less stringent than the 1998 one.
More data $\Longrightarrow$ less information!

\bigskip

\begin{tabular}{cc}
\parbox{0.45\textwidth}{\footnotesize
\includegraphics[width=0.45\textwidth]{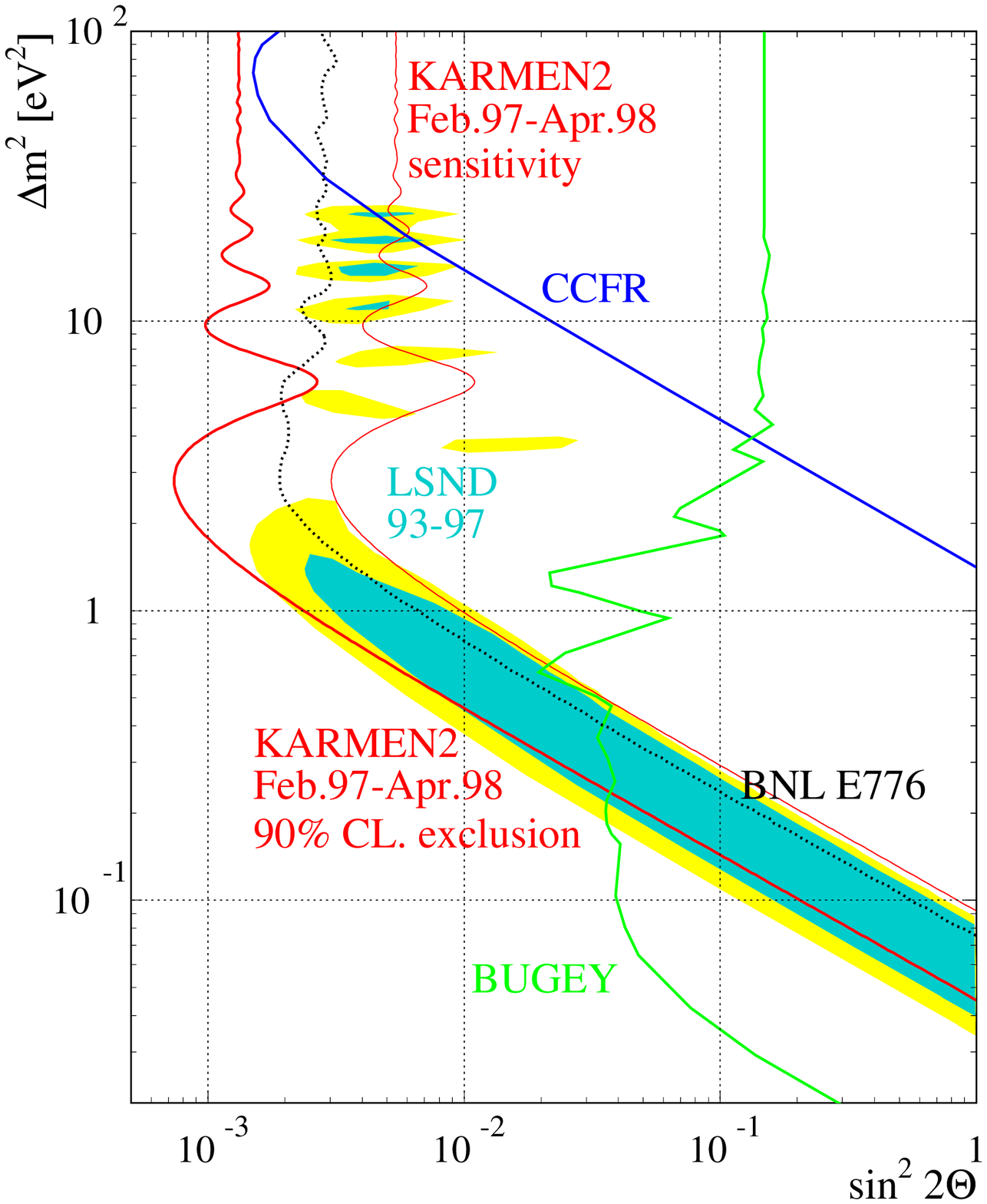}
KARMEN 1998 exclusion limit and sensitivity \cite{KARMEN-nu98}.
}
&
\parbox{0.45\textwidth}{\footnotesize
\includegraphics[bb=50 150 497 660, width=0.45\textwidth]{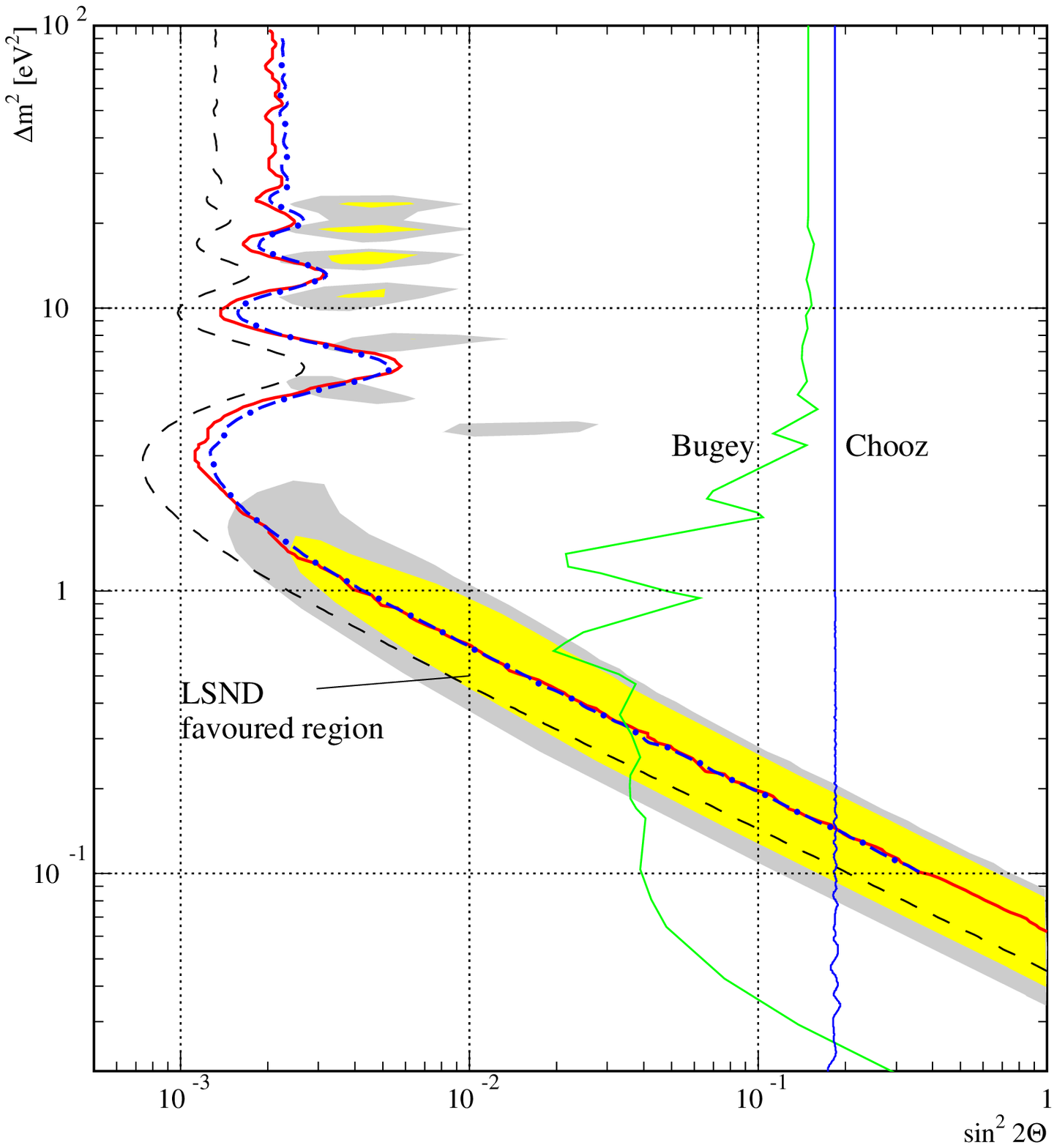}
KARMEN 1999 exclusion limit and sensitivity \cite{KARMEN-Moriond-99}.
Solid line: 1999 limit.
Dash-Dotted line: 1999 sensitivity.
Dashed line: 1998 limit.
}
\end{tabular}

\newpage 

{\large\red{$\nRightarrow$}}
Other infamous examples:

{\large\red{$\nRightarrow$}}
The 1999 limit on neutrinoless
double-beta decay obtained in the \textbf{Heidelberg-Moscow}
experiment \cite{Heidelberg-Moscow-99} obtained with the Unified Approach
was much more stringent than the sensitivity of the experiment.
That is why now they do not use the Unified Approach any more
\cite{Heidelberg-Moscow-01}! (Got burned!)

{\large\red{$\nRightarrow$}}
The present upper limit on $\nu_\mu\to\nu_\tau$
neutrino oscillations obtained in the NOMAD experiment \cite{NOMAD-00}
is stronger than the one obtained in the CHORUS experiment \cite{CHORUS-01}
not because the NOMAD experiment has a better sensitivity than
the CHORUS experiment
(see discussion in \cite{CHORUS-01}), but
because the NOMAD collaboration uses the Unified Approach,
which gives unphysically stringent upper bounds
when the number of observed events is smaller than the expected background.

\newpage 

{\large\red{$\lightning$}}
The UA and similar methods
\cite{Giunti-bo-99,Ciampolillo-98,Mandelkern-Schultz-99}
are appropriate in order to get allowed regions
even in the presence of an unlikely statistical fluctuation of the data,
such that the data are very unlikely for any value of the parameters.

However, the physical reliability of such allowed regions
is highly questionable.

{\large\red{$\lightning$}}
If there is no statistical fluctuation of data,
the UA and the CG methods
are \underline{equivalent}.

{\large\red{$\lightning$}}
From the value of the GoF
(see Table at pag.~\pageref{tab-gof})
one can see that
\textbf{there is no unlikely statistical fluctuation
in solar neutrino data}
in the case of $\nu_e\to\nu_{\mu,\tau}$ oscillations
and in the case of the analysis of the rates
in terms of
$\nu_e\to\nu_{s}$ oscillations.

On the other hand,
if the solar neutrino problem is due to
$\nu_e\to\nu_{s}$ oscillations,
there is an unlikely statistical fluctuation
of the shape of the energy spectrum
and the global analysis of solar $\nu$ data
with the CG method is unreliable.

{\large\red{$\lightning$}}
Therefore,
the CG method
(that is computationally much easier than the UA method)
can be applied to the analisis of solar $\nu$ data
in terms of $\nu_e\to\nu_{\mu,\tau}$ oscillations
and to the analysis of the rates of solar neutrino experiments
in terms of $\nu_e\to\nu_s$ oscillations.

\newpage 

\includegraphics[bb=120 505 370 780, width=0.49\textwidth]{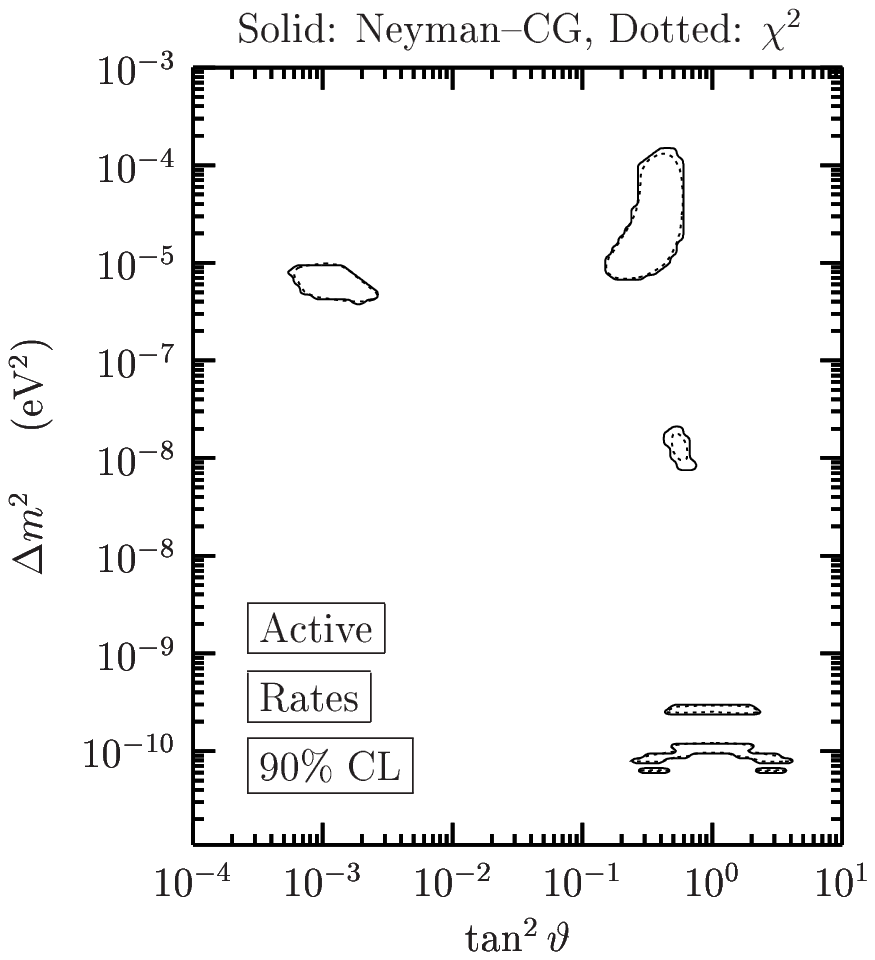}
\includegraphics[bb=120 505 370 780, width=0.49\textwidth]{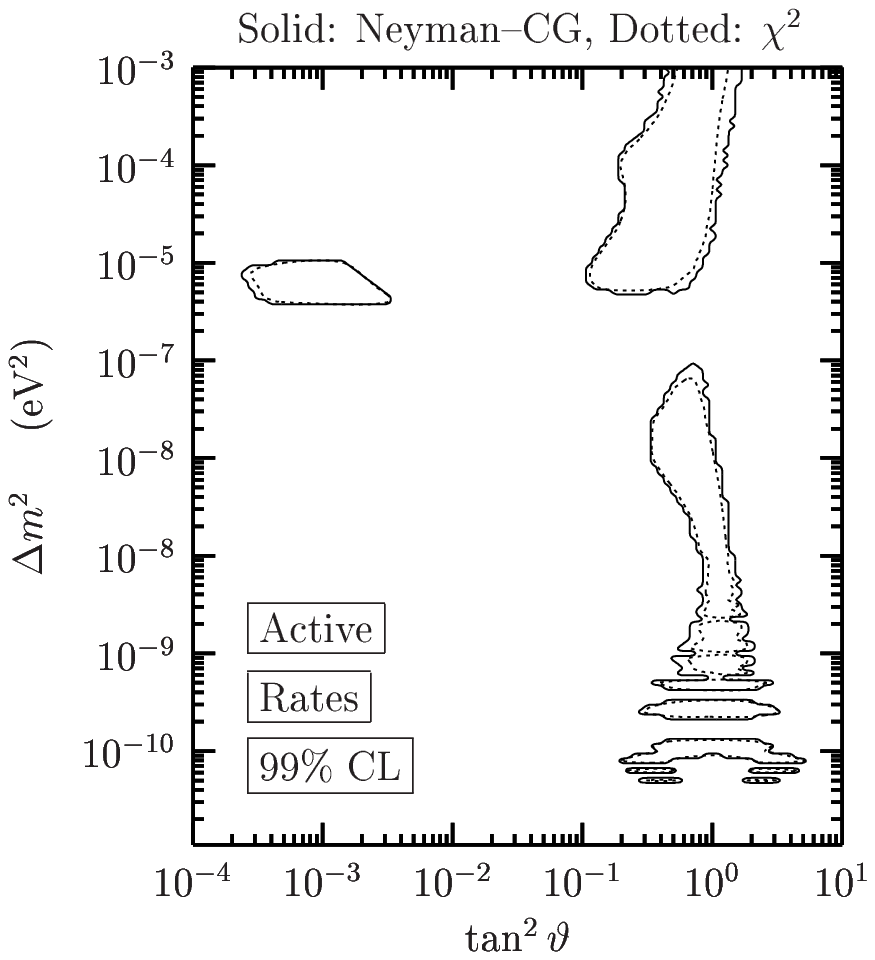}

\includegraphics[bb=120 505 370 780, width=0.49\textwidth]{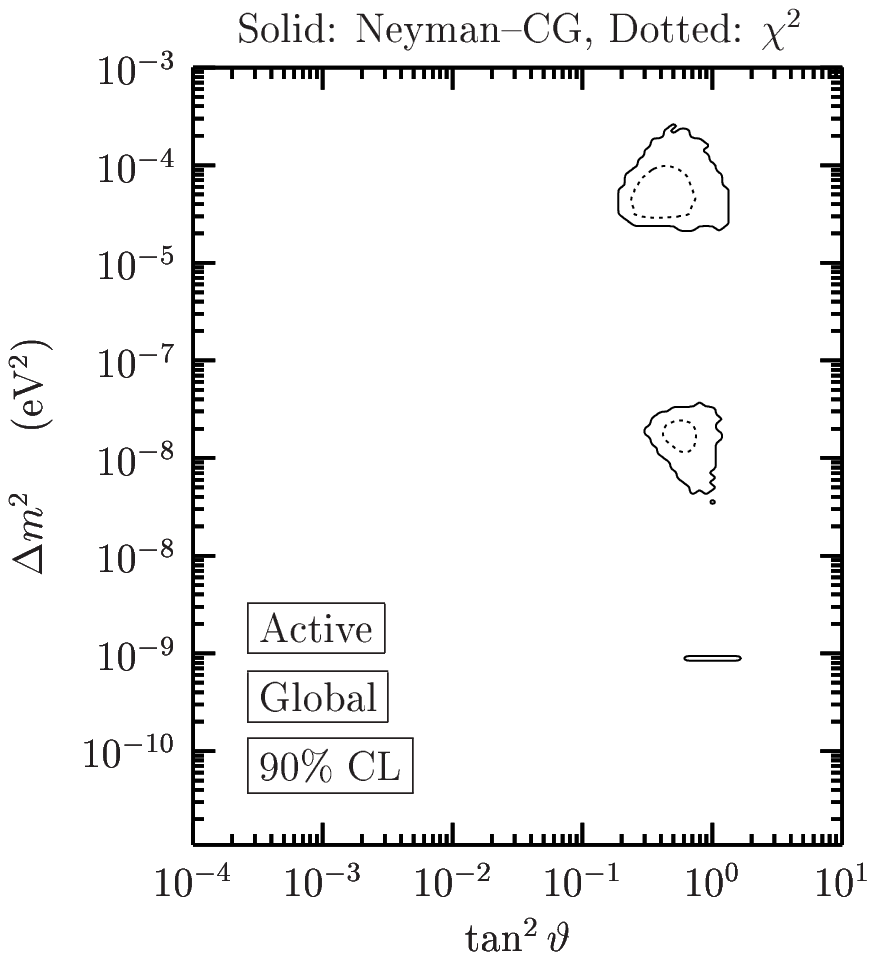}
\includegraphics[bb=120 505 370 780, width=0.49\textwidth]{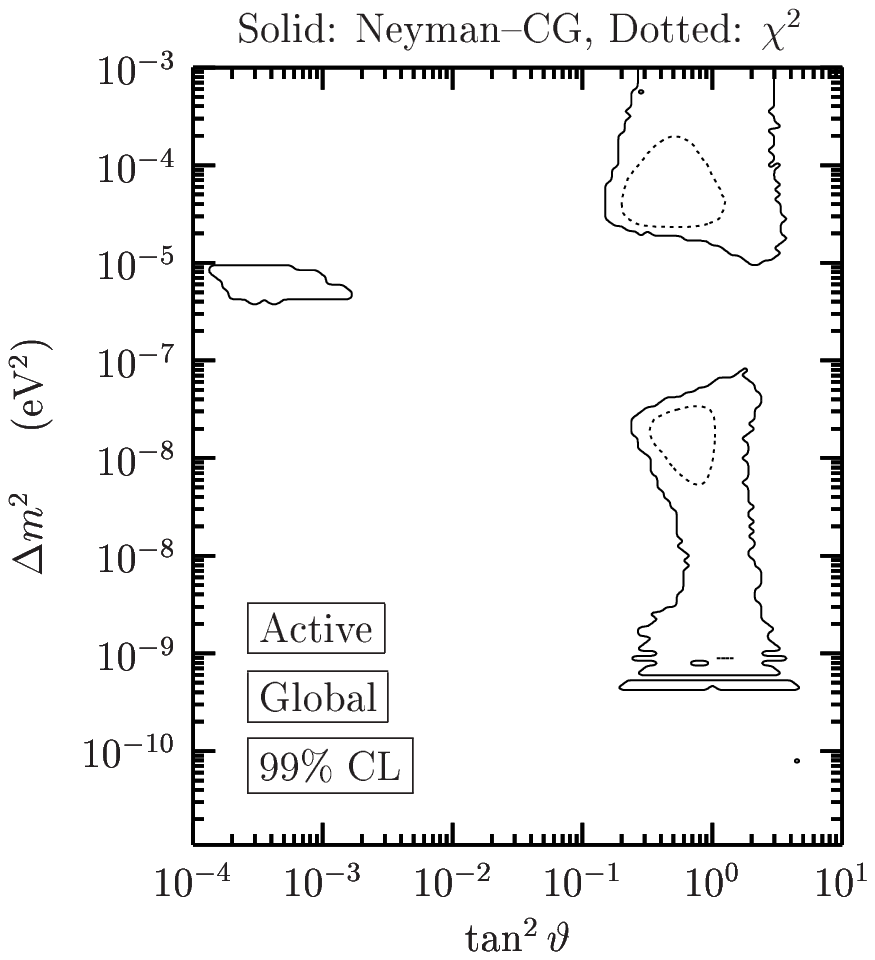}

\newpage 

\includegraphics[bb=120 505 370 780, width=0.49\textwidth]{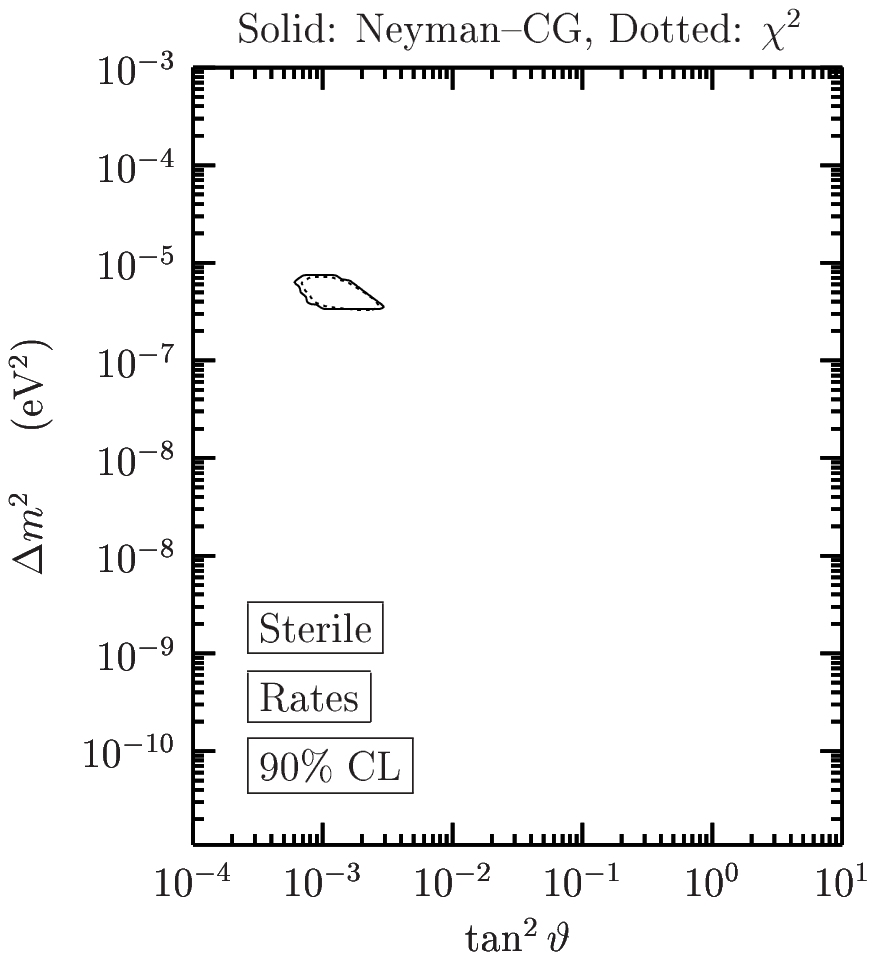}
\includegraphics[bb=120 505 370 780, width=0.49\textwidth]{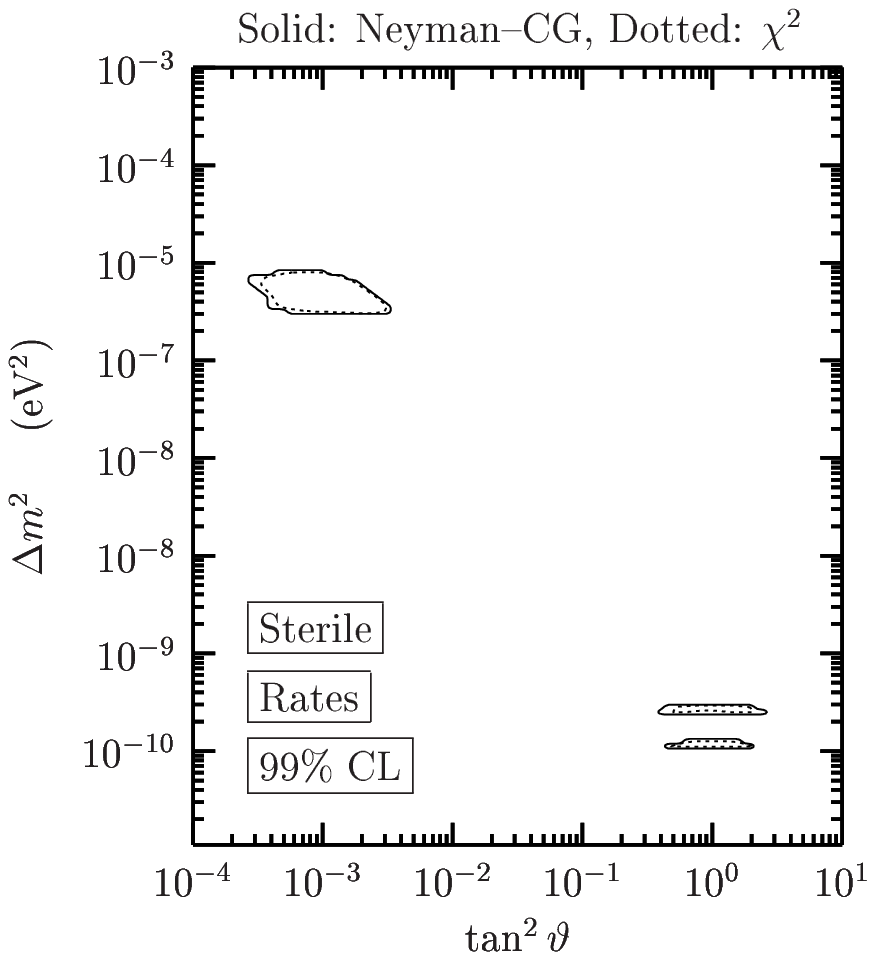}

\newpage 

\begin{equation*}
\text{\Large\textbf{\porpora{Bayesian Allowed Regions}}}
\end{equation*}

{\large\red{$\Rightarrow$}}
Bayesian Theory
allows to calculate the \textbf{improvement of knowledge
as a consequence of experimental measurements}
(see \cite{D'Agostini-99}).

{\large\red{$\Rightarrow$}}
This is how our mind works and how science improves.
Therefore,
\textbf{Bayesian Theory is the natural statistical tool for scientists}.

{\large\red{$\blacktriangleright$}}
Bayesian probability density function of
$\tan^2\vartheta$ and $\Delta{m}^2$
after measurement of rates $R^{\mathrm{(exp)}}_{j}$:
\begin{equation*}
p(\tan^2\!\theta,\Delta{m}^2|R^{\mathrm{(exp)}}_{j})
\propto
\mathcal{L}(R^{\mathrm{(exp)}}_{j}|\tan^2\!\theta,\Delta{m}^2)
p(\tan^2\!\theta,\Delta{m}^2)
\end{equation*}
$p(\tan^2\!\theta,\Delta{m}^2)$
=
prior probability density function

{\large\red{$\lightning$}}
\underline{Prior knowledge on $\tan^2\vartheta$ and $\Delta{m}^2$}:
All values are allowed,
but we know that solar $\nu$ data
are sensitive to different orders of magnitude
of $\tan^2\vartheta$ and $\Delta{m}^2$
through different mechanisms.
\begin{equation*}
{\red{\Downarrow}}
\end{equation*}
\begin{equation*}
\text{Flat prior in the
$\log(\tan^2\!\vartheta)$--$\log(\Delta{m}^2)$ plane}
\end{equation*}
\begin{equation*}
{\red{\Downarrow}}
\end{equation*}
\begin{equation*}
\fbox{$ \displaystyle
\begin{array}{l}
p(\tan^2\!\theta,\Delta{m}^2|R^{\mathrm{(exp)}}_{j})
=
\\ \displaystyle
=
\frac
{\mathcal{L}(R^{\mathrm{(exp)}}_{j}|\tan^2\!\theta,\Delta{m}^2)}
{\int \mathcal{L}(R^{\mathrm{(exp)}}_{j}|\tan^2\!\theta,\Delta{m}^2)
\,
\mathrm{d}\!\log(\tan^2\!\vartheta)
\,
\mathrm{d}\!\log(\Delta{m}^2)
}
\end{array}
$}
\end{equation*}

\newpage 

\includegraphics[bb=120 505 370 780, width=0.49\textwidth]{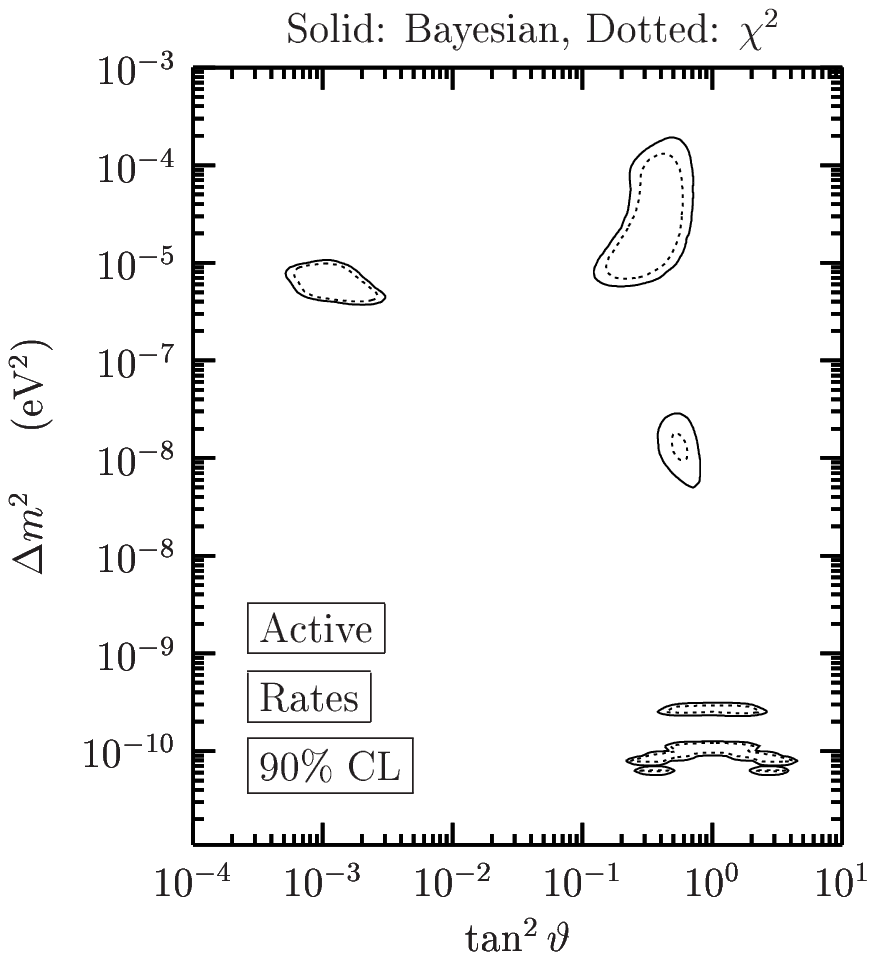}
\includegraphics[bb=120 505 370 780, width=0.49\textwidth]{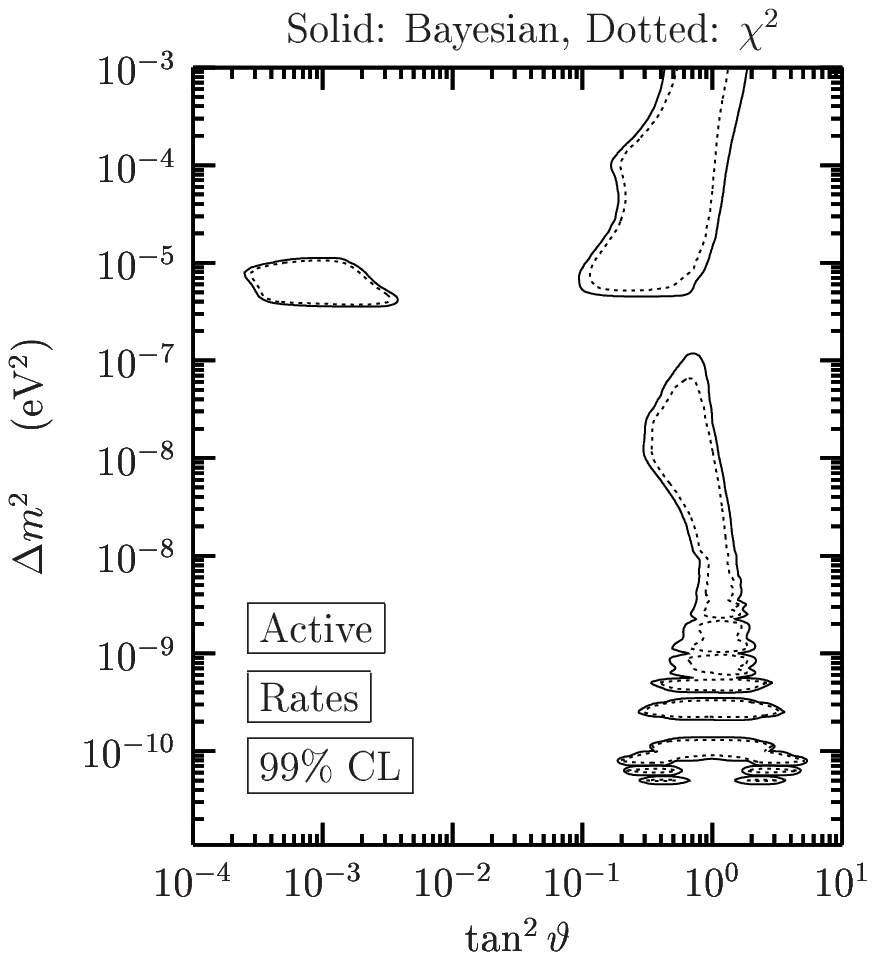}

\includegraphics[bb=120 505 370 780, width=0.49\textwidth]{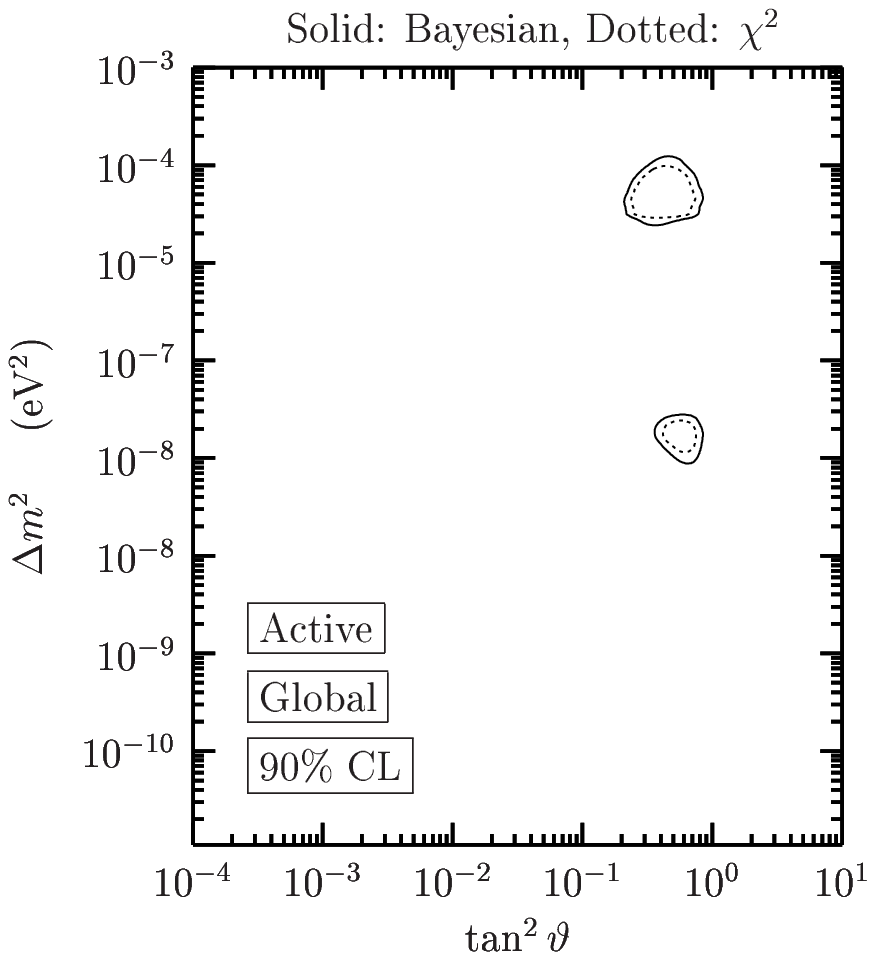}
\includegraphics[bb=120 505 370 780, width=0.49\textwidth]{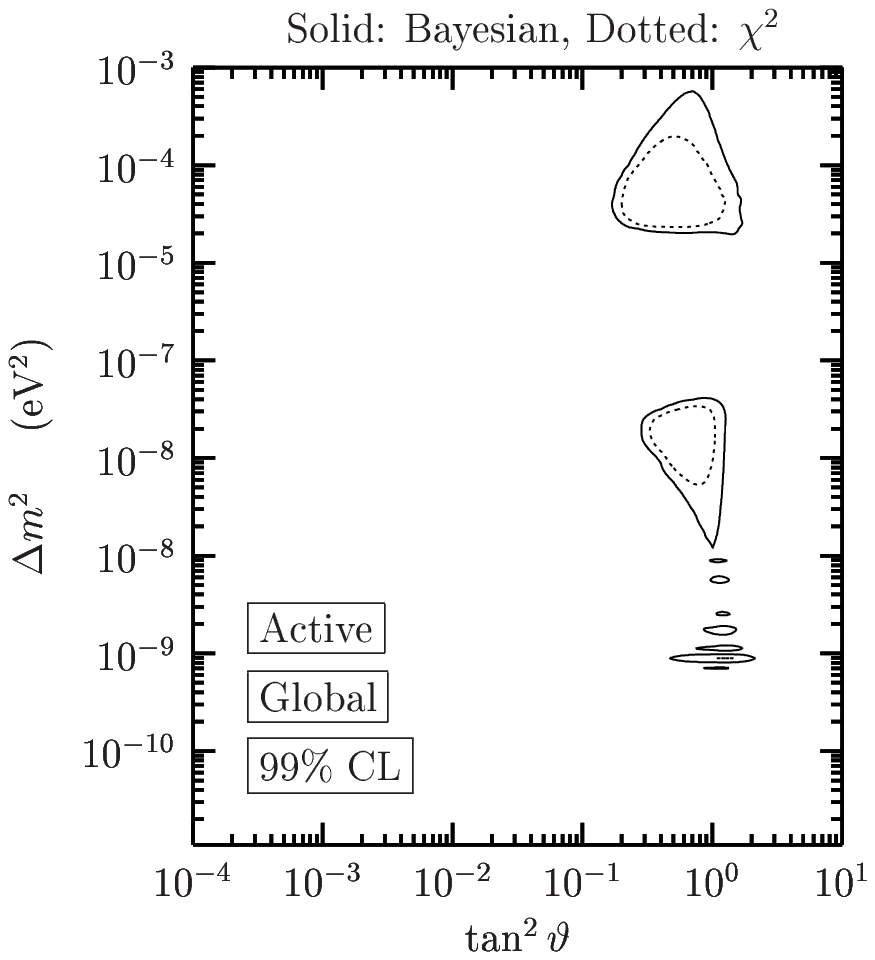}

\newpage 

\includegraphics[bb=120 505 370 780, width=0.49\textwidth]{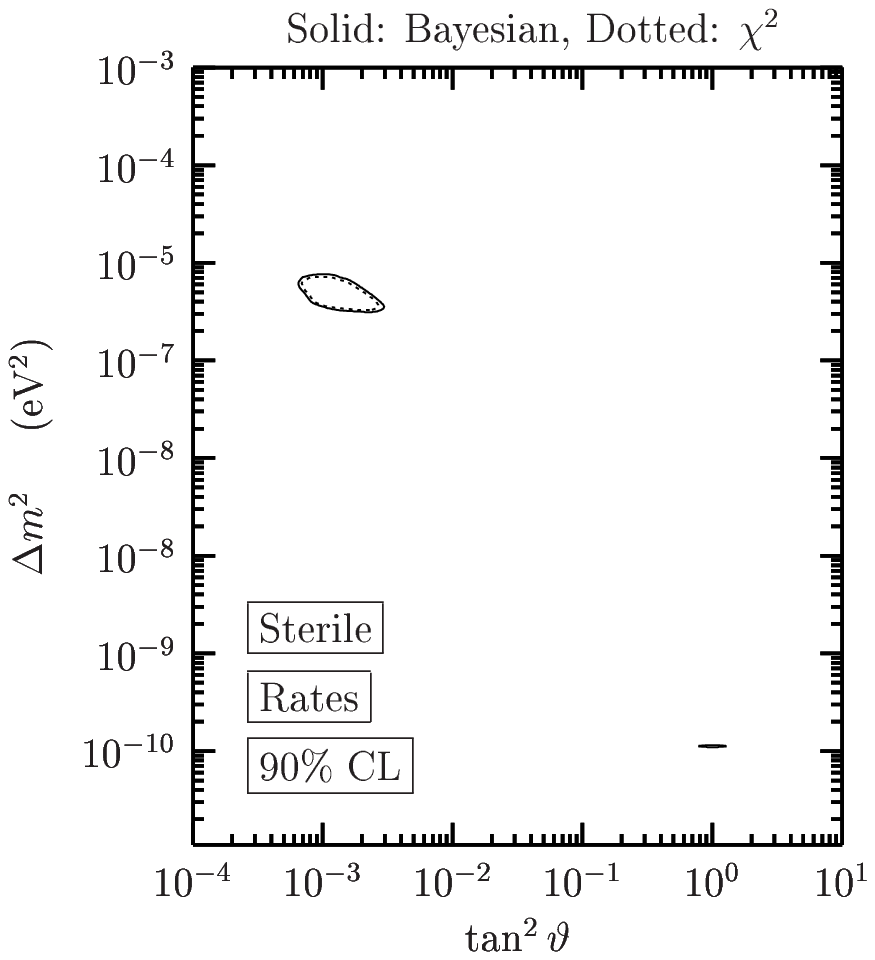}
\includegraphics[bb=120 505 370 780, width=0.49\textwidth]{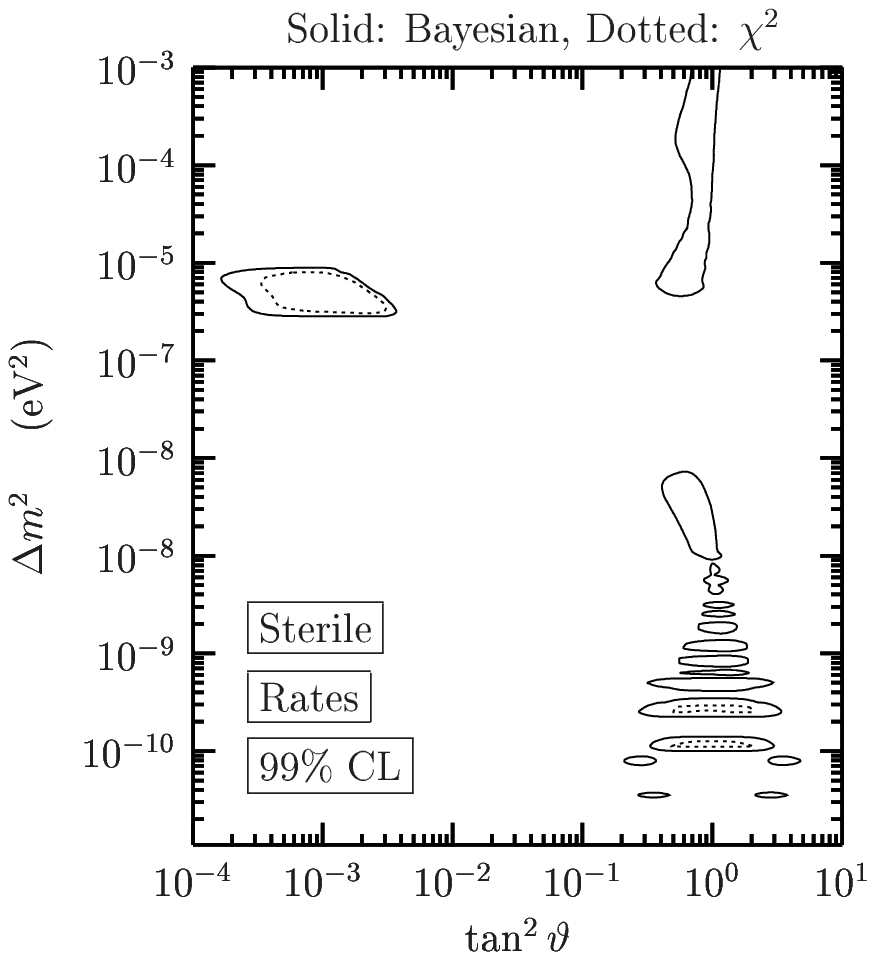}

\includegraphics[bb=120 505 370 780, width=0.49\textwidth]{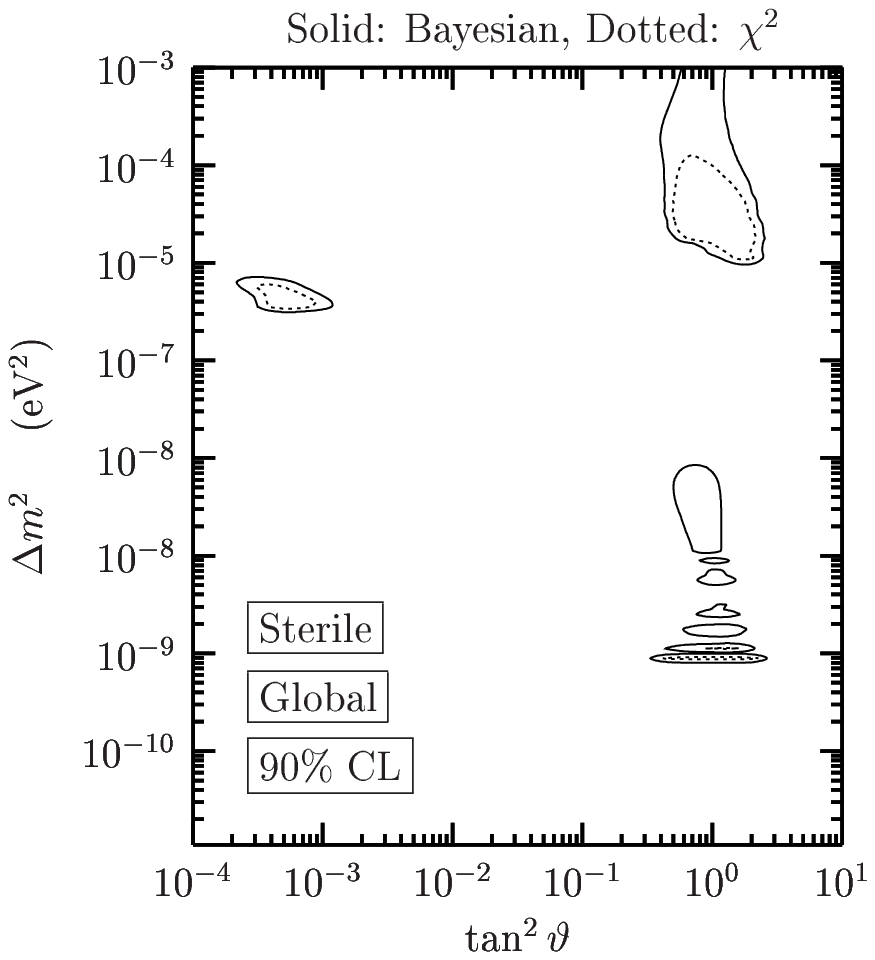}
\includegraphics[bb=120 505 370 780, width=0.49\textwidth]{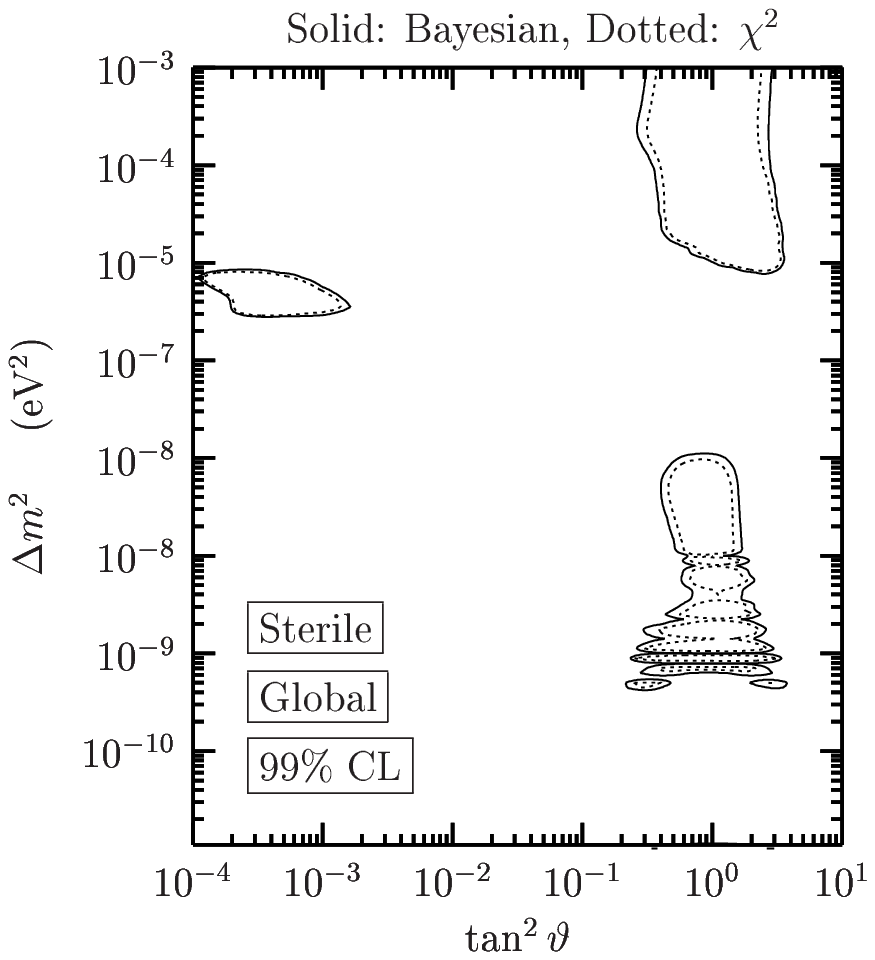}

\newpage 

\normalsize

\providecommand{\href}[2]{#2}\begingroup\raggedright\endgroup

\end{document}